\documentclass[aps,twocolumn,groupedaddress,showpacs]{revtex4}

\newcommand{\M}{{\cal M}}
\newcommand{\T}{{\cal T}}
\newcommand{\E}{{\cal E}}

\begin{document}

\title{Qubit authentication}

\author{Marcos Curty}
\email[]{mcurty@com.uvigo.es}

\author{David J. Santos}
\email[]{dsantos@com.uvigo.es}

\author{Esther P\'erez}
\email[]{estherperez@edu.xunta.es}
\affiliation{ETSIT, Universidad de Vigo, Campus Universitario s/n,
E-36200 Vigo (Spain)}

\author{Priscila Garc\'{\i}a-Fern\'andez}
\email[]{pgarcia@foton0.iem.csic.es}
\affiliation{Instituto de \'Optica, CSIC, Serrano, 123, E-28006 Madrid (Spain)}

\date{\today}

\begin{abstract}
Secure communication requires message authentication. In this paper we
address the problem of how to authenticate quantum information sent
through a quantum channel between two communicating parties with the
minimum amount of resources.  Specifically, our objective is to determine
whether one elementary quantum message (a qubit) can be authenticated
with a key of minimum length. We show that, unlike the case of
classical-message quantum authentication, this is not possible.  
\end{abstract}


\maketitle

\section{Introduction}

Cryptography deals about communication in the presence of adversaries
\cite{RIVEST_1990}, and one of its fundamental goals is to provide
message authentication (also called data-origin authentication)
\cite{MENEZES_1996}. This is a process whereby a party is corroborated
as the original source of some specified data created, transmitted,
or stored at some time in the past. To assure message authentication,
one then must have the ability to detect any data manipulation by
unauthorized parties.  This is particularly important in public-key
cryptography, in which users must be confident about the authenticity
of the public-keys of the partners involved in the communication.

Classical cryptography provides data-origin authentication by
means of two general techniques: message authentication codes (MACs)
\cite{WEGMAN_1981}, and digital-signature schemes \cite{DIFFIE_1976}. In
both cases, the authentication process is specified by two 
algorithms: an encoding, or tagging algorithm (possibly stochastic),
and a decoding or verification algorithm. When the sender (Alice)
wishes to send a certified message to a recipient (Bob), she generates,
employing the encoding algorithm, an authentication tag or a signature
(in both cases a function of the message and a secret encoding-key) and
appends it to the message before actually sending it. Notice that this
tagged-message may be sent in the clear; the authentication problem is,
therefore, very different to the one associated to encryption since no
secrecy is necessary for secure message authentication. On the reception
side, Bob verifies the authenticity of the message by means of the
specified decoding procedure, which depends on the message, the tag,
and a decoding-key. This algorithm returns a bit indicating when Bob
must regard the message as authentic, and accept it as coming from Alice,
and when he must discard it. The basic requirement is that the tags,
which are produced by the encoding algorithm, be accepted as valid by
the verification algorithm when a decoding-key corresponding to the
encoding-key is used on the tagging procedure. When a message
authentication scheme fulfills this requisite it is said that it
provides perfect deterministic decoding.

Choosing between MACs and digital-signature schemes depends on the
communication context. Typically, MACs are employed when the receiver is
predetermined at the time of message transmission, and the decoding-key
he owns, which can be equal to the encoding-key, is secret.  In signature
schemes, on the contrary, the decoding-key is public, and therefore known
also to the adversary (Eve); this fact guarantees that anybody can verify
the authenticity of data (universal verification), and that Alice cannot
later repudiate having signed the message, since no one but she owns
the encoding-key.

Although several information-theoretic secure MACs have been proposed
(see, e.g., \cite{WEGMAN_1981}), the security provided by signature
schemes depends on unproven assumptions related to the intractability of
certain difficult mathematical problems, such as the prime-factorization
of large numbers \cite{RIVEST_1978} or the discrete-logarithm computation
\cite{KRAVITZ_1993}. Unfortunately, if a quantum computer is ever built,
Shor's quantum algorithms \cite{SHOR_1997} could break classical signature
schemes easily, i.e., in polynomial time.

The use of quantum resources in user authentication has been proposed
before in QKD scenarios \cite{LJUNGGREN_2000,ZENG_2000}. More recently,
several proposals of general message authentication using quantum
resources have been made. Gottesman and Chuang \cite{gottesman_2001}
have proposed a quantum version of classical digital-signature
schemes. Their technique requires limited circulation of the public
key, but it allows unconditionally secure authentication of classical
messages. A disadvantage of this protocol, as stated by the authors,
is that it requires several non-reusable key-bits for each message-bit
signed.  Leung \cite{LEUNG_2000}, in another recent paper, also studies
quantum message authentication. Her protocol requires a two-way classical
channel between partners, thus making the overall security dependent on
the security of that channel. In a previous article \cite{curty_2001_b}
some of us proposed a class of quantum authentication protocols that
allow secure authentication of classical binary messages with one bit as
the authentication key.  This improves the efficiency of classical MACs,
that require, for the authentication of binary messages, at least two-bit
keys. In this paper we adapt this class of protocols to the authentication
of quantum messages (qubits), and study its security.  We show that,
using the amount of quantum resources needed to authenticate one-bit
classical messages, the intrinsic nature of quantum information makes it
impossible to detect every unauthorized quantum data manipulation by a
potential adversary, thus making qubit authentication not possible.

The paper is organized as follows. In Section~\ref{PROTOCOL} we
describe a general class of one-qubit message authentication protocols.
This class can be seen as a generalization of the one presented
in \cite{curty_2001_b} to authenticate classical binary messages. In
Section~\ref{SECURITY} we analyze the security of these protocols against
several attacks.  First, we analyze the no-message attack, in which
the sender has not initiated the transmission (there is no message in
the channel), and Eve attempts to prepare a fake quantum message with
the intention of passing Bob's verification test.  Then, we study more
subtle attacks, those in which Eve has access to what is transmitted.
After studying all these attacks we show that it is not possible to keep
the failure probability of the authentication protocol below one.

\section{Quantum message authentication}
\label{PROTOCOL}

Suppose Alice needs to send a certified quantum message to Bob.
The goal is to make Bob confident about the authenticity of the message
and sender. For the sake of simplicity, let us assume that the state to
send certified is an arbitrary qubit described by the density operator
$\rho_\M$ operating on some two-dimensional message space $\M$.  This
qubit may be locally generated by her, or she may be acting as a relay
station between two other parties.  In order to certify this message,
Alice follows the standard procedure in classical authentication: She
appends a tag to the message in such a way that the recipient, Bob,
may verify the tag and so convince himself about the identity of the
message originator. The difference with respect to the classical case is
that now the tag is also a quantum state, given by the density operator
$\rho_\T$, of some tag space $\T$.  Therefore, the quantum tagged-message
is described by the operator $\rho_\E=\rho_\M\otimes\rho_\T$ that acts
on the state space $\E=\M\otimes\T$.  Since our objective is to determine
whether authentication is possible with the minimum amount of resources,
we shall regard $\T$ as a two-dimensional space.  Although this might
seem a strong restriction, it can be shown that having a bigger tag does
not improve the security of the protocol, at least when dealing with an
ideal, error-free quantum channel, as is the case considered here. In
a noisy channel, a bigger tag space would certainly be useful in the
detection of errors in the channel that might alter the message, in the
way Quantum Error Correcting Codes (QECC) usually work.  We shall return
to this point with more detail later, at the end of Section III. Finally,
since we are interested in perfect deterministic decoding, valid and
invalid tags must belong, respectively, to orthogonal subspaces in
$\T$, but, since one cannot find orthogonal mixed states
in a two-dimensional space ($\T$), $\rho_\T$ must be a pure state,
namely $\rho_\T=|0\rangle\langle 0|_\T$.

No authentication is possible without a previously shared secret
between the two communicating parties. This key, which may have
been exchanged directly or by means of a trusted third party (a
certification authority), can be a classical or a quantum one.  In our
proposal we shall assume that Alice and Bob share a maximally entangled
quantum state as their secret authentication key. For instance, each
of the parties could own one qubit of a publicly-known singlet state
$|\psi\rangle_{AB}=\frac{1}{\sqrt{2}}(|01\rangle_{AB}-|10\rangle_{AB})$.
It can be argued that, in a realistic scenario, dealing with classical
keys would be more advantageous. In fact, our protocol can equally operate
with one-bit classical keys. However, we prefer the use of quantum keys
for their better key-management properties ---no copying of the key
remains undetected if extra quantum key is included for this check.

The main difference between authenticating a classical message and a
quantum one resides in the nature of $\rho_\M$. If $\rho_\M$ is a quantum
message, it belongs to the continuous space of density operators acting
on $\M$. Since Alice may act just as a relay station, we shall assume
that $\rho_\M$ is unknown to her. In the classical case, there are only
two classical messages to send, `0' or `1'. To encode them unambiguously
Alice and Bob need to agree on a particular orthonormal basis of $\M$,
$\{|0\rangle_\M,|1\rangle_\M\}$, so $\rho_\M$ would be restricted to
be either $|0\rangle\langle 0|_\M$ or $|1\rangle\langle 1|_\M$. This
decision can be made openly, adding no more secrecy between the parties.

According to the notation introduced above, the state of the global system
(key+tagged-message) is given by

\begin{equation}
\rho_{AB\E}
=|\psi\rangle\langle\psi|_{AB}\otimes\rho_\E
=|\psi\rangle\langle\psi|_{AB}\otimes\rho_\M
\otimes|0\rangle\langle 0|_\T.
\end{equation}

\noindent Next Alice performs, on her part of $|\psi\rangle_{AB}$ and
on the tagged-message, an encoding operation $E_{A\E}$.  We may write
this unitary operation in the general form:

\begin{equation}
E_{A\E}=|0\rangle\langle 0|_A \otimes 1_{\cal E}+
        |1\rangle\langle 1|_A \otimes U_{\cal E},
\label{ENCODING}
\end{equation}

\noindent where $U_{\cal E}$ is some arbitrary unitary quantum
operation on $\E$.  Basically, the action of $E_{A\E}$ is equivalent
to the selection, triggered by the state of a one-bit key, of one
operator from an arbitrary pair of publicly-known unitary ones (in
our case, $1_\E$ and $U_\E$).  Once this operator is selected, it is
applied to $\rho_\E$ before sending it through the quantum channel.
The reason to enforce the unitarity of these two operators is that
it allows Bob to easily undo their action. But this is not the only
way in which perfect deterministic decoding can be achieved. As shown
in \cite{CAVES_1997,NIELSEN_1998,CAVES_1999}, more general quantum
operations, such as trace-preserving completely-positive (TPCP) maps,
can, under certain circumstances, be reversed, and, therefore, used in
$E_{A\E}$ instead of $1_\E$ and $U_\E$. However, for simplicity in the
formalism we limit ourselves to the case of unitary operators.

If we denote by $\rho^e_{AB\E}$ the density operator describing
the state of the global system after the encoding operation, i.e.,
$\rho^e_{AB\E}=E_{A\E} \rho_{AB\E}E_{A\E}^\dagger$, then the state of the
tagged-message that Alice sends to Bob through the quantum channel between
them is given by $\rho^e_\E=\mbox{tr}_{AB}\left(\rho^e_{AB\E}\right)$,
where $\rho^e_{AB\E}$, making use of (\ref{ENCODING}), can be written as

\begin{eqnarray}
\label{RHO_E_ABE}
\rho_{AB\E}^e&=&\frac{1}{2}\left(|01\rangle\langle 01|_{AB}\otimes \rho_\E+
|10\rangle\langle 10|_{AB}\otimes U_\E \rho_\E U_\E^\dagger\right.\nonumber\\ 
&-&\left.
|01\rangle\langle 10|_{AB}\otimes\rho_\E U_\E^\dagger-
|10\rangle\langle 01|_{AB}\otimes U_\E\rho_\E\right).
\end{eqnarray}

\noindent From this expression, it is easy to obtain the tagged-message
in the channel:

\begin{equation}
\rho^e_\E=\frac{1}{2}(\rho_\E+U_{\cal E}\rho_\E U_{\cal E}^{\dag}).
\label{prot1}
\end{equation}

On the reception side, Bob decodes the information sent by
Alice performing the unitary decoding operation

\begin{equation}
D_{B\E}=|0\rangle\langle0|_B U_{\cal E}^\dagger+|1\rangle\langle 1|_B 1_{\cal E}
\label{DECO_OP}
\end{equation}

\noindent on his qubit of the singlet and the tagged-message
received: $\rho^d_{AB\E}=D_{B\E}\rho^e_{AB\E}D_{B\E}^\dagger$.
Using (\ref{RHO_E_ABE}) and (\ref{DECO_OP}) in this equation one
can easily calculate the decoded tagged-message as
$\rho^d_\E=\mbox{tr}_{AB}(\rho^d_{AB\E})=\rho_\E$.  Finally, Bob
verifies the state of the tag: He performs the orthogonal
measurement $\{|0\rangle\langle 0|_\T,|1\rangle\langle 1|_\T\}$
over the tag-portion of $\rho^d_\E$, where $|1\rangle_\T$
is the state in $\T$ orthogonal to the tag $|0\rangle_\T$. If the
result of such a measurement is $|0\rangle_\T$, Bob should assume
that no tampering has taken place, and therefore extract the
quantum message sent to him;  otherwise, he rejects the message
received.


\section{Security analysis}
\label{SECURITY}

In the previous section we have claimed that the proposed tag-based
quantum authentication protocol provides perfect deterministic decoding.
This means that the protocol would fail only if Bob accepted a message
as an authenticated one when that is not the case (due to the unnoticed
action of Eve). When dealing with forgery strategies we must consider
two main types of attacks: The no-message attack, and the message
attack \cite{MENEZES_1996}. The first one is the simpler: Before
any message is sent by Alice to Bob, Eve attempts to prepare a quantum
state that passes the decoding algorithm. The message attack is more
subtle and severe: Eve could access the authentic messages transmitted,
and try to produce a forged message based on the information gained. In
the following discussion we shall show how, unlike the case of
authentication of classical messages \cite{curty_2001_b}, when dealing
with a one-qubit message it is impossible to select a unitary operation
$U_{\cal E}$ that makes all Eve's possible attacks unsuccessful.
We find necessary conditions for the probability of successful forgery
by a no-message attack  (in section IIIA), and by a measurement attack
(in section IIIB-1), to be less than one. Then in section IIIB-2 we show
that under these conditions, there exists another attack, consisting of
a unitary operator applied to the message and tag system, that succeeds
with probability one.

\subsection{No-message attack}
\label{SUBSEC_FORGERY}

Suppose that Eve prepares some quantum state $\rho^E_\E \in \E$
and sends it to Bob trying to impersonate Alice.  When Bob receives
this quantum message he cannot know that it comes from a forger,
so he follows the procedure explained in the previous section:
He performs the decoding operation $D_{B\E}$ and then an orthogonal
measurement $\{|0\rangle\langle 0|_\T,|1\rangle\langle 1|_\T\}$
over the tag space. Before all this takes place, Bob `sees', as the
encoded global state, the density operator
$\rho_{AB\E}^e=|\psi\rangle\langle\psi|_{AB}\otimes \rho^E_\E$.
After Bob's decoding, the tagged-message is given by

\begin{equation}
\rho^d_\E=\mbox{tr}_{AB}\left(\rho_{AB\E}^d\right)
=\frac{1}{2}\left(\rho^E_\E+U_\E^\dagger\rho^E_\E U_\E\right),
\end{equation}

\noindent where $\rho_{AB\E}^d=D_{B\E}\rho_{AB\E}^e D_{B\E}^\dagger$.
As we have seen, Bob rejects the message if the result of his orthogonal
measurement on $\T$ is $|1\rangle_\T$; therefore, the
probability $P_f$ that Eve deceives Bob is:

\begin{equation}
P_f=\langle 0|\mbox{tr}_\M
\left[
\frac{1}{2}
\left(
\rho^E_\E +U_\E^\dagger \rho^E_\E U_\E
\right)
\right]
|0\rangle_\T.
\end{equation}

\noindent This quantity depends both on Eve's strategy ---her selection
of $\rho_\E^E$--- and on the quantum operation $U_{\cal E}$. She will
succeed with probability one if she chooses a $\rho_\E^E$ that satisfies

\begin{equation}
\mbox{tr}_\M\left(\rho^E_\E \right)=|0\rangle\langle 0|_\T,
\label{PRIMERA}
\end{equation}

\noindent and

\begin{equation}
\mbox{tr}_\M
\left(
U_\E^\dagger \rho^E_\E U_\E
\right)=|0\rangle\langle 0|_\T.
\label{SEGUNDA}
\end{equation}

\noindent In order to check whether Eve succeeds, let us first
consider the case in which she prepares a pure state
$\rho_\E^E=|\phi\rangle\langle \phi|_\E$. The conditions (\ref{PRIMERA})
and (\ref{SEGUNDA}) above can then be rewritten as

\begin{equation}
\mbox{tr}_\M\left(|\phi\rangle\langle
\phi|_\E\right)=|0\rangle\langle 0|_\T,
\label{PRIMERA_BIS}
\end{equation}

\noindent and

\begin{equation}
\mbox{tr}_\M\left(U_\E^\dagger|\phi\rangle\langle
\phi|_\E U_\E\right)=|0\rangle\langle 0|_\T,
\label{SEGUNDA_BIS}
\end{equation}

\noindent respectively. From equation (\ref{PRIMERA_BIS}) we conclude that
Eve should choose $|\phi\rangle_\E=|\psi\rangle_\M\otimes |0\rangle_\T$,
with $|\psi\rangle_\M$ any pure state in $\M$. When this result is used in
(\ref{SEGUNDA_BIS}), then

\begin{equation}
U_\E^\dagger \left(|\psi\rangle_\M\otimes
|0\rangle_\T\right)= |\omega\rangle_\M\otimes |0\rangle_\T
\label{LIMBO_3}
\end{equation}

\noindent should also be satisfied, where $|\omega \rangle_\M$ is some
pure state in $\M$.

Without loss of generality, let us write $U_\E$ in the form

\begin{equation}
U_\E=\sum_{i,j=0}^1 U_{ij}\otimes |i\rangle\langle j|_\T,
\label{NUEVA}
\end{equation}

\noindent where the $U_{ij}$, with $i,j=0,1$, act on
$\M$. With this notation, condition (\ref{LIMBO_3}) requires
$U_{01}^\dagger|\psi\rangle_\M=0$, i.e., the operator $U_{01}$ must be
singular.  Therefore, if Alice and Bob select a unitary operation $U_\E$
such that $U_{01}$ is nonsingular, then $P_f<1$ for any pure state
prepared by Eve.

What if Eve prepares a general mixed state $\rho_\E^E$? Any mixed state in
$\E$ can be spectrally decomposed as $\rho_\E^E=\sum_{i=0}^3\lambda_i
|\phi_i\rangle \langle \phi_i|_\E$, where
$0\leq\lambda_i\leq 1$, $\sum_{i=0}^3\lambda_i=1$, and $\langle
\phi_i|\phi_j\rangle_\E=\delta_{ij},\,\, i,j=0,\cdots,3$. When
this decomposition of $\rho_\E^E$ is used in (\ref{PRIMERA}) and
(\ref{SEGUNDA}), these equations are transformed into a problem equivalent
to the one posed by equations (\ref{PRIMERA_BIS}) and (\ref{SEGUNDA_BIS}).
Therefore, also in this more general case we can assure the result
$P_f<1$.  In fact, we can further show that, with the appropriate
selection of $U_\E$, $P_f$ can be made at most $1/2$.  As we have seen,
$P_f$ can be written as

\begin{equation}
P_f=\langle 0|\mbox{tr}_\M\left({\rho_\E^d}\right)|0\rangle_\T,
\end{equation}

\noindent with ${\rho_\E^d}=\left(\rho_\E^E+U_\E^\dagger\rho_\E^E
U_\E\right)/2$. Defining the projector $P=|00\rangle\langle
00|_\E+|10\rangle\langle 10|_\E$, $P_f=\mbox{tr}\left({\rho_\E^d}
P\right)$. Using the properties of the trace operator,

\begin{equation}
P_f=\mbox{tr}\left(\rho_\E^E Q\right)/2,
\end{equation}

\noindent where $Q=U_\E P U_\E^\dagger+P$ is a positive operator known
to Eve, and with maximum eigenvalue $\lambda_{max}\geq 1$. Therefore,
the maximizing $\rho_\E^E$ is any eigenvector corresponding to
$\lambda_{max}$, and thus $P_f=\lambda_{max}/2$.  Finally, it is easy
to see (see, e.g., \cite{HORN_1985}) that choosing $U_\E$ such that it
takes $P$ to its orthogonal complement makes $\lambda_{max}=1$, and
then, as predicted, $P_f=1/2$.

\subsection{Message attack}

This is a more subtle and severe family of attacks. Instead of
directly forging a quantum message and send it to Bob, Eve could
wait for Alice's message and manipulate it. Proceeding this way
she tries to convert authentic messages into others with high
probability of passing Bob's test.

In order to simplify the analysis, and without loss of generality, we
shall distinguish between two types of message attacks. In the first one,
Eve tries to extract information, by means of the appropriate measurement
of the message in the channel, that allows her to prepare a different
message that Bob regards as authentic.  In the second class of attacks,
Eve does not care about the current message in the channel. Instead,
based on the knowledge of all the public aspects of the quantum
authentication scheme used, she determines a quantum operation and
applies it to any data sent by Alice. This quantum operation can be
described by a TPCP map.

\subsubsection{Measurement}

\label{APARTADO_MEDIDA}

According to equation (\ref{prot1}), the information to which Eve
has access is

\begin{equation}\label{prot1_bis}
\rho^e_\E=
\frac{1}{2}[\rho_\M\otimes |0\rangle\langle 0|_\T+U_{\cal E}
(\rho_\M\otimes |0\rangle\langle 0|_\T) U_{\cal E}^{\dag}].
\end{equation}

\noindent Since Eve knows how the protocol works, she could get
information about the key if, performing the appropriate measurement on
the channel, she could perfectly distinguish between the two terms on
the right-hand side of (\ref{prot1_bis}). If Eve managed to achieve it,
she would collapse the state of Alice and Bob shared-key in a known
unentangled pure quantum state, so she could throw away the authentic
message and prepare and send to Bob a new one that would pass his
test. Because Eve does not know which quantum message, $\rho_\M$,
has been sent, the only way to discern between the two terms is by
means of their tag-portions.  Therefore, in order to make this strategy
successful, the states $|0\rangle \langle 0|_\T$ and $\mbox{tr}_\M \left[
U_\E\left(\rho_\M\otimes |0\rangle\langle 0|_\T\right)U_\E^\dagger \right]$
must be perfectly distinguishable, i.e.,

\begin{equation}
\langle 0|
\mbox{tr}_\M \left[ U_\E\left(\rho_\M\otimes
|0\rangle\langle 0|_\T\right)U_\E^\dagger \right]
|0\rangle_\T=0.
\label{CONDMEASURE}
\end{equation}

\noindent Making use of the spectral decomposition of $\rho_\M$ (see
our reasoning in Section \ref{SUBSEC_FORGERY}), the requirement
above can be alternatively written, without loss of generality, as

\begin{equation}
\mbox{tr}_\M
\left(
U_\E|\phi\rangle\langle \phi|_\M \otimes |0\rangle\langle 0|_\T
U_\E^\dagger
\right)=|1\rangle\langle 1|_\T,
\label{LIMBO_2}
\end{equation}

\noindent where $|\phi\rangle_\M$ is any state in $\M$.  Equation
(\ref{LIMBO_2}) is satisfied when $U_\E \left(|\phi\rangle_\M\otimes
|0\rangle_\T\right)=|\omega\rangle_\M\otimes |1\rangle_\T$, where
$|\omega\rangle_\M$ is some state in $\M$. Following a procedure parallel
to the one employed in Section \ref{SUBSEC_FORGERY}, we obtain the
requirement that $U_{00}$ must be singular.  Therefore, if Alice and
Bob select an operation $U_\E$ such that $U_{00}$ is nonsingular, Eve
cannot infer from her measurement the necessary information about the key.

\subsubsection{TPCP map}

Consider that Alice sends to Bob a quantum tagged-message $\rho_\E$. If
no eavesdropping takes place, the state in the channel is given by
(\ref{prot1}). But, assume now that Eve has the power to perform an
arbitrary TPCP map, $\$$, on the tagged-message sent. Eve wants to choose
$\$ $ such that the decoding procedure performed by Bob on the resulting
state led to any state not equal to $\rho_\M$ while the tag remains
in the state  $|0\rangle_\T$.

The global state resulting from Bob's decoding operation after Eve's
TPCP mapping is $\rho_{AB\E}^d= D_{B\E} \rho_{AB\E}^E D_{B\E}^\dagger$,
where $\rho_{AB\E}^E=\$ \left(\rho_{AB\E}^e\right)$, with $\rho_{AB\E}^e$
given by (\ref{RHO_E_ABE}). The tagged-message decoded by Bob is

\begin{equation}
\rho^d_\E=\mbox{tr}_{AB}\left(
\rho_{AB\E}^d\right)=\frac{1}{2}\left[
\$ \left(\rho_\E\right)+U_\E^\dagger\$\left(U_\E \rho_\E
U_\E^\dagger\right) U_\E\right].
\end{equation}

\noindent The probability of Bob's accepting the message as a valid one
is $P_{TPCP}=\langle 0| \mbox{tr}_\M\left(\rho^d_\E\right)|0\rangle_\T$.
Clearly, this probability is one if and only if
$\mbox{tr}_\M(\rho^d_\E)=|0\rangle\langle 0|_\T$, which means
that the conditions

\begin{eqnarray}
\mbox{tr}_\M\left[\$\left(\rho_\E\right)\right]&=&|0\rangle\langle
0|_\T,\label{LA_20}\\
\mbox{tr}_\M\left[U_\E^\dagger
\$\left(U_\E \rho_\E U_\E^\dagger \right)
 U_\E\right]&=&
|0\rangle\langle 0|_\T,
\end{eqnarray}

\noindent must be simultaneously satisfied. In order to check whether
Eve succeeds, let us first consider the most simple case of TPCP,
that in which she performs a unitary operation $A_\E$. Using again
the spectral decomposition of $\rho_\M$ (see, again, our reasoning in
Section \ref{SUBSEC_FORGERY}), the conditions above can be alternatively
written as

\begin{eqnarray}\label{TPCP1}
\mbox{tr}_\M\left[A_\E\left(|\phi\rangle\langle\phi|_\M\otimes|0\rangle
\langle{}0|_\T\right)A_\E^{\dagger}\right]&=&|0\rangle\langle
0|_\T,\\ \mbox{tr}_\M\left[B_\E \label{TPCP2}
\left(|\phi\rangle\langle\phi|_\M\otimes|0\rangle\langle{}0|_\T
\right) B_\E^{\dagger} \right]&=& |0\rangle\langle 0|_\T,
\end{eqnarray}

\noindent where $|\phi\rangle_\M$ is any state in $\M$, and
$B_\E=U_\E^\dagger A_\E U_\E$.  Then, it is straightforward to see
that, in order to fulfill (\ref{TPCP1}) and (\ref{TPCP2}), Eve needs to
choose a unitary operation such that

\begin{eqnarray}
A_\E&=&\sum_{i=0}^1 A_{ii} \otimes |i\rangle\langle{}i|_\T,
\label{AA}\\
B_\E&=&\sum_{i=0}^1
B_{ii}\otimes |i\rangle\langle{}i|_\T,
\label{BB}
\end{eqnarray}

\noindent where the $A_{ii}$ and $B_{ii}$, $i=0,1$, are some unitary
operations on the state space $\M$.  Since $B_\E$ and $A_\E$ are related
by $B_\E=U_\E^\dagger A_\E U_\E$, we can arrange equations (\ref{AA})
and (\ref{BB}) in matrix form and write

\begin{equation}
\left(
\begin{array}{cc}
U_{00}^\dagger & U_{10}^\dagger\\
U_{01}^\dagger & U_{11}^\dagger\\
\end{array}
\right)
\cdot
\left(
\begin{array}{cc}
A_{00} & 0\\
0 & A_{11}\\
\end{array}
\right)
\cdot
\left(
\begin{array}{cc}
U_{00} & U_{01}\\
U_{10} & U_{11}\\
\end{array}
\right)
=
\left(
\begin{array}{cc}
B_{00} & 0\\
0& B_{11}\\
\end{array}
\right).
\label{ECUACION_MATRICIAL}
\end{equation}

\noindent According to our formulation of the problem, Eve would succeed
in her attack if, given the unitary operator $U_\E$, she could always find
two unitary operators $A_\E$ and $B_\E$ such that $U_\E B_\E=A_\E U_\E$. Of
course, the trivial solution $A_\E=B_\E=I_\E$ (Eve does not modify
the state of the tagged-message) is discarded.

In Section \ref{SUBSEC_FORGERY} we have shown that Alice and Bob can avoid
Eve's forgering of messages if $U_{01}$ is nonsingular. In a similar
way, in Section \ref{APARTADO_MEDIDA} we have seen that Alice and Bob
can prevent Eve from gaining critical information about the key by way
of measurement if $U_{00}$ is also nonsingular. If we assume now that
these conditions hold, i.e., that $U_{00}$ and $U_{01}$ are nonsingular,
then it can be shown, using the unitarity of $U_\E$, that $U_{10}$ and
$U_{11}$ are then also nonsingular. Under these conditions, as we show
in Appendix \ref{DEMO_CANADIENSE}, the equation $U_\E B_\E=A_\E U_\E$
has always (for any $U_\E$) the prescripted solution, and Eve can always
be successful in her attack. This unconditional success of Eve's attack
makes unnecessary to consider the more general case of a TPCP map.

\subsection{Discussion on the structure of the tag space}

Let us now briefly explain why increasing the dimension of the tag space
does not affect any of the arguments considered in this section.

As seen by Eve, the key shared by Alice and Bob controls whether the
quantum state in the channel belongs to the two-dimensional subspace
spanned by the vectors $\{|0\rangle_{\M} |0\rangle_{\T} ,|1\rangle_{\M}
|0\rangle_{\T} \}$ or to the one spanned by $\{ U_{\E}
\left(|0\rangle_{\M} |0\rangle_{\T} \right),U_{\E}\left(|1\rangle_{\M}
|0\rangle_{\T} \right)\}$, so there are two alternative coding
subspaces for the valid quantum messages.  All the conditions the
unitary operator $U_{\E}$ has to fulfill in order to avoid Eve's
attacks can be geometrically interpreted as conditions on the relative
position between these two code subspaces of the four-dimensional space
${\E}$. For instance, in order to avoid the success of the no-message
attack, as stated by the conditions (\ref{PRIMERA})-(\ref{SEGUNDA}),
the subspaces must span $\E$. On the other hand, in order to avoid the
measurement attack, condition (\ref{CONDMEASURE}), both subspaces must
be non-orthogonal.  As for the unitary attack, we have seen that Eve can
always fulfill conditions (\ref{AA})-(\ref{BB}), i.e. she can always find
a unitary operator acting on ${\E}$ such that it leaves invariant both
subspaces at the same time, independently of their relative position,
and rotate the vectors within in a non trivial way.

If the dimension of $\T$ is increased, the dimension of the space ${\E}$
also increases (say to $N$, with $N>4$), but if the valid tag state is
still some particular state $|{\bf 0}\rangle_{\T} \equiv |0 \cdots
0\rangle_{\T}$, the dimension of the two alternative code subspaces
will remain equal to two.  Thus we are just embedding the problem in
a larger space, but not changing its intrinsic complexity: In
a $N$-dimensional space Alice and Bob would still be able to
choose two two-dimensional code subspaces neither linearly dependent
nor orthogonal, and Eve would still find a unitary nontrivial operator
in $U(N)$ whose restriction to the particular four-dimensional space
containing the two codes leaves both codes invariant.

One may also consider what would happen if we allowed the tag state
to be a mixed, $\rho_{\T}$, rather than a pure state. According to our
protocol, in order to have perfect deterministic decoding, the tag should
belong to any subspace of the tag space (so that Alice an Bob could
perfectly distinguish between valid or invalid messages). When measuring
a mixed state one can obtain any state belonging to the subspace the
mixed state has support on (with a certain probability, given by the
spectral decomposition).  Consequently, when Bob verifies the tag, he
will consider as valid any result giving a state inside that subspace, and
will discard the message if the result lays in the orthogonal complement.
Since the dimension of the two distinct code subspaces increases, things
are slightly more complicated when the tag is not pure. Let $n<N$ be
the dimension of the valid code subspaces, and let us see whether we
can generalize conditions (\ref{PRIMERA})-(\ref{SEGUNDA}) and
(\ref{CONDMEASURE})-(\ref{LIMBO_2}).

Generalization of conditions (\ref{PRIMERA})-(\ref{SEGUNDA}) above
is achieved if and only if, for some state $\rho^E_{\E}$, the states
$\mbox{tr}_\M\left(\rho^E_\E \right)$ and $\mbox{tr}_\M \left(U_\E^\dagger
\rho^E_\E U_\E \right)$ belong to the valid tag subspace. But, as argued
above, this can be avoided by Alice and Bob choosing an operator $U_\E$
such that the block of $U^\dagger_\E$ transforming states from the valid
tag subspace to the invalid one (the analogous to $U_{01}^\dagger$ in
(\ref{NUEVA})) is nonsingular. Note that if $n<N/2$ this block is not a
square operator, and nonsingular means with trivial kernel. Note also
that choosing $n>N/2$ (a subspace of valid messages bigger than the
invalid one) is insecure, since there would always be vectors of the
valid subspace transformed into the null state by that block, and so
Eve would be successful in the no-message attack with probability one.

Generalization of equations (\ref{CONDMEASURE})-(\ref{LIMBO_2}),
related to the measurement attack, is achieved if and only if the state
$\mbox{tr}_\M \left[ U_\E\left(\rho_\M\otimes \rho_\T\right)U_\E^\dagger
\right]$  belongs to the invalid tag subspace. But again this can be
avoided choosing $U_\E$ such that the equivalent to block $U_{00}$
(now a $n^2$-dimensional block operator, transforming states within the
valid tag subspace) is nonsingular.

But the crucial fact is that Eve can still fulfill conditions
(\ref{AA})-(\ref{BB}), i.e., there is always a nontrivial unitary
operator leaving invariant the two code subspaces. To see this,
note that the generalization of equations (\ref{LA_20})--(\ref{TPCP2})
is obtained by simply imposing that the states in the left-hand side
of those equations belong to the valid tag subspace. Then equations
(\ref{AA})--(\ref{ECUACION_MATRICIAL}) would have the same form,
apart from the fact that now $A_{00}$, $B_{00}$ and $U_{00}$ would
represent $n^2$-dimensional operators, $A_{11}$, $B_{11}$ and $U_{11}$
$(N-n)^2$-dimensional ones, $U_{01}$ a $n \times (N-n)$ operator,
and $U_{10}$ a $(N-n) \times n$ one.  The solutions for $A_{00}$ and
$A_{11}$ given in the appendix equally hold if $n=N/2$ (equations
(\ref{AP_UNA})-(\ref{AP_OTRA}) are exactly the same, all the blocks are
square operators, and we can invert them). If $n<N/2$ then we are just
embedding a $2n<N$ problem into a larger $N$-dimensional space, and thus
the solution of the intrinsic problem still exists.

The fact that increasing the dimension of $\T$ does not improve the
robustness of the protocol may surprise the reader familiar with Quantum
Error Correction Codes (QECC). In these codes more qubits of tag are
added to protect against more errors. But note that the nature of the
errors considered is statistical, i.e. they are randomly generated
by the noise in the channel. In QECC the efficiency of the correction
capability lays on the assumption that statistical errors on a large
number of qubits are less likely than those on a small number. The
errors considered here, arbitrary unitary actions on the space $\E$,
do not belong to this category.

\section{Conclusion}
\label{CONCLUSIONES}

Providing message authentication is one of the main goals of communication
security. Classical message-authentication methods can be combined with
quantum teleportation to make the authentication of quantum information
possible. However, it is not yet clear whether this procedure is
optimal in the resources it requires. In this paper we study the
authentication of elementary quantum messages (qubits) using minimum-size
keys. Specifically, we have generalized a previous class of quantum
authentication protocols \cite{curty_2001_b} to the case of one-qubit
messages, and studied its security against a forger with quantum power
and full access to the channel between the communicating parties. The
main result of this study is that, unlike classical binary messages,
the intrinsic nature of quantum information makes the authentication
of one qubit using a minimal key impossible, i.e. it is not possible
to keep the failure probability of the qubit authentication protocol
below one. We are currently investigating the minimum amount of quantum
resources needed to authenticate a qubit, and the use of more general
quantum operations (TPCP maps) in the encoding and decoding actions. Our
results will be published elsewhere.

\section*{Acknowledgments}

The authors acknowledge Howard Barnum for sharing with them his early
work on quantum message authentication codes, and Ad\'an Cabello, Debbie
Leung, and an anonymous referee for their insightful comments. We
gratefully thank some members of the Applied Mathematics Department at
the University of Vigo, particularly, I. Area and A. Mart\'{\i}n, as
well as R. Israel, from the University of British Columbia, for their
assistance with Matrix Analysis. This work was partially supported by
Xunta de Galicia (Spain, grant n.\ PGIDT00PXI322060PR) and the Spanish
Government (grant n.\ TIC2001-3217).


\appendix

\section{}
\label{DEMO_CANADIENSE}

In this appendix we show that, given a unitary operator $U_\E$, acting
on a four-dimensional space $\E=\M \otimes \T$, and  with the
$2\times 2$ blocks of its matrix representation nonsingular, one (Eve)
can always find two unitary, block-diagonal operators , $A_\E$ and $B_\E$,
such that $U_\E ^\dagger A_\E U_\E=B_\E$.

The explicit expressions for the operators in their block form can be
written as:

\begin{eqnarray}
U_\E= \sum_{i=0}^1 \sum_{j=0}^1
U_{ij}\otimes |i\rangle\langle j|_\T,  \\
A_\E=\sum_{i=0}^1 A_{ii} \otimes |i\rangle\langle{}i|_\T,
\label{AA2} \\
 B_\E=\sum_{i=0}^1
B_{ii}\otimes |i\rangle\langle{}i|_\T.
\label{BB2}
\end{eqnarray}

\noindent From the equality of the  blocks resulting from multiplying
by $|0\rangle\langle 1|_\T$ and $|1\rangle\langle 0|_\T$ both sides
of $U_\E ^\dagger A_\E U_\E=B_\E$, we obtain the equations:

\begin{eqnarray}\label{OFF1} U_{10}^\dagger  A_{11}   U_{11}  +
U_{00}^\dagger  A_{00}   U_{01} = 0, \label{AP_UNA}
 \\ \label{OFF2}  U_{11}^\dagger
A_{11}   U_{10}  + U_{01}^\dagger  A_{00}   U_{00} =0.\label{AP_OTRA}
\end{eqnarray}

\noindent Since all the $U_{ij}$ blocks are invertible, we have the following
two expressions for $A_{11}$:

\begin{eqnarray}
 A_{11}  &=& - U_{10}^{\dagger^{-1}} U_{00}^\dagger  A_{00}   U_{01}
 U_{11}^{-1}  \nonumber \\  &=& - U_{11}^{\dagger^{-1}} U_{01}^\dagger
 A_{00}   U_{00}  U_{10}^{-1}     \label{2A11}
\end{eqnarray}

\noindent The second equality in this equation implies: $ A_{00}   U_{01}
U_{11}^{-1}U_{10} U_{00}^{-1} = U_{00}^{\dagger^{-1}} U_{10}^\dagger
U_{11}^{\dagger^{-1}} U_{01}^\dagger  A_{00}$, that is, $A_{00}G=G^\dagger
A_{00}$, with $G= U_{01} U_{11}^{-1}U_{10} U_{00}^{-1}$. Using the
following relation between the $U_{ij}$ blocks, derived from the unitarity
of $U_\E$,

\begin{equation}
 U_{00}^\dagger  U_{01}  + U_{10}^\dagger  U_{11} = 0,  \label{unit1}
\end{equation}

\noindent we can write $G=-U_{00}^{{-1}^\dagger}U_{10}^\dagger U_{10}
U_{00}^{-1}$, so $G$ is hermitian, and, therefore, $A_{00}$ and $G$
commute. Any hermitian operator commutes with some unitary operator that
is not a scalar multiple of the identity, so this guarantees $A_{00}$
exists. Given $A_{00}$, we can obtain $A_{11}$ from  (\ref{2A11}). Now
it remains to be shown that such $A_{11}$ is also unitary. Computing
the adjoint of the first expression in (\ref{2A11}) and multiplying it
by the second, we obtain:

\begin{equation}
A_{11}^\dagger A_{11}
=U_{11}^{\dagger^{-1}}  U_{01}^\dagger  A_{00}^\dagger   U_{00}
U_{10}^{-1} U_{11}^{\dagger^{-1}} U_{01}^\dagger  A_{00}   U_{00}
U_{10}^{-1}.
\end{equation}

\noindent From (\ref{unit1}), $U_{11}^{\dagger}U_{10}=- U_{01}^{\dagger}
U_{00}$. Applying this result to the equation above leads to

\begin{eqnarray}
A_{11}^\dagger A_{11} &=&
- U_{11}^{{-1}^\dagger}  U_{01}^\dagger  A_{00}^\dagger  A_{00}  U_{00}
U_{10}^{-1}\\ \nonumber
&=&  - U_{11}^{{-1}^\dagger} U_{01}^\dagger U_{00}  U_{10}^{-1}
=U_{11}^{{-1}^\dagger}  U_{11} U_{10}  U_{10}^{-1} = I,
\end{eqnarray}

\noindent and, therefore, $A_{11}$ is unitary.

As for $B_{00}$ and $B_{11}$, they can be obtained, in terms of $A_{00}$
and $A_{11}$, from the equality between the  blocks of $U_\E ^\dagger A_\E
U_\E=B_\E$, when both of its sides are multiplied by $|0\rangle\langle
0|_\T$ and $|1\rangle\langle 1|_\T$.  Their unitarity is proven in a
similar way.

\end{document}